\documentclass[a4paper,12pt]{article}

\usepackage{url}
\usepackage{epsfig}
\usepackage{amsmath}
\usepackage{amssymb}
\usepackage{amsfonts}
\usepackage{bbm}
\usepackage[footnotesize]{caption}
\usepackage{graphicx}
\usepackage{mathrsfs}
\usepackage[font=scriptsize]{subcaption}
\usepackage{xcolor}
\usepackage{comment}
\usepackage{mathtools}
\usepackage{braket}
\usepackage{cite}
\usepackage{hyperref}
\usepackage{multirow}
\usepackage{relsize}
\usepackage{fullpage}
\usepackage{bbm}
\usepackage[T1]{fontenc}
\usepackage{pifont}
\usepackage{makecell}

\allowdisplaybreaks

\definecolor{darkpink}{RGB}{219, 48, 122}

\newcommand{\Dmq}{\Delta m^2}

\newcommand{\eVq}{\ensuremath{\text{eV}^2}}
\renewcommand{\Re}{\mathop{\mathrm{Re}}}
\renewcommand{\Im}{\mathop{\mathrm{Im}}}
\usepackage{graphicx} 
\usepackage{mathtools}

\newcommand{\newc}{\newcommand}
\newc{\be}{\begin{equation}}
\newc{\ee}{\end{equation}}
\newc{\bea}{\begin{eqnarray}}
\newc{\eea}{\end{eqnarray}}
\newc{\simlt}{~\mbox{\smaller\(\lesssim\)}~}
\newc{\simgt}{~\mbox{\smaller\(\gtrsim\)}~}

\begin{document}

\begin{titlepage}

\begin{center}
{\bf\Large  
Modular flavour symmetry and orbifolds
} \\[12mm]
Francisco~J.~de~Anda$^{\ddagger}$%
\footnote{E-mail: \texttt{fran@tepaits.mx}},
Stephen~F.~King$^{\diamond}$%
\footnote{E-mail: \texttt{king@soton.ac.uk}}
\\[-2mm]

\end{center}
\vspace*{0.50cm}
\centerline{$^{\ddagger}$ \it
Tepatitl{\'a}n's Institute for Theoretical Studies, C.P. 47600, Jalisco, M{\'e}xico,}
\centerline{\it
Dual CP Institute of High Energy Physics, C.P. 28045, Colima, M\'exico.}
\vspace*{0.2cm}
\centerline{$^{\diamond}$ \it
School of Physics and Astronomy, University of Southampton,}
\centerline{\it
SO17 1BJ Southampton, United Kingdom.}
\vspace*{1.20cm}

\begin{abstract}
{\noindent 
We develop a bottom-up approach to flavour models which combine modular symmetry with orbifold constructions.
We first consider a 6d orbifold $\mathbb{T}^2/\mathbb{Z}_N$, with a single torus defined by one complex coordinate 
$z$ and a single modulus field $\tau$, playing the role of a flavon transforming under a finite modular symmetry.
We then consider 10d orbifolds with three factorizable tori, each defined by one complex coordinate $z_i$ and involving the three moduli fields $\tau_1, \tau_2, \tau_3$ transforming under three finite modular groups.
Assuming supersymmetry, consistent with the holomorphicity requirement, 
we consider all 10d orbifolds of the form $(\mathbb{T}^2)^3/(\mathbb{Z}_N\times\mathbb{Z}_M)$,
and list those which have fixed values of the moduli fields (up to an integer).
The key advantage of such 10d orbifold models over 4d models is that the values of the moduli are not completely free but are 
constrained by geometry and symmetry.
To illustrate the approach we discuss a 10d modular seesaw model with $S_4^3$ modular symmetry 
based on $(\mathbb{T}^2)^3/(\mathbb{Z}_4\times\mathbb{Z}_2)$
where $\tau_1=i,\ \tau_2=i+2$ are 
constrained by the orbifold, while $\tau_3=\omega$ is determined by imposing a further remnant $S_4$ flavour symmetry,
leading to a highly predictive example in the class CSD$(n)$ with $n=1-\sqrt{6}$.
}
\end{abstract}

\end{titlepage}

\section{Introduction}

The Standard Model (SM), despite its many successes, does not account for the origin of neutrino mass nor the
quark and lepton family replication, and gives no insight into the fermion masses and mixing parameters.
One approach is to introduce a family symmetry which may be a finite discrete or continuous, gauged or global, Abelian or non-Abelian.
Large lepton mixing has motivated studies of non-Abelian finite discrete groups such as $A_4,S_4,A_5$ (for reviews see e.g.~\cite{King:2013eh,King:2017guk}). However such family symmetries must eventually be 
spontaneously broken by new Higgs fields called flavons, and it turns out that the vacuum alignment of such flavon fields plays
a crucial role determining the physical predictions of such models. 

Another interesting class of symmetries arise from the modular group $SL(2,\mathbb{Z})$, which is the group of $2\times 2 $
matrices with integer elements, the kind you first learn about in high school with positive or negative elements, but with unit determinant.
Geometrically, such a group is the symmetry of a torus, which essentially has a flat geometry in two dimensions (when it is cut open) and the symmetry corresponds to the discrete coordinate transformations which leave the torus invariant, in other words the different choices of two dimensional lattice vectors describing the same torus. The two dimensional space may be conveniently associated with 
the real and imaginary directions of the complex plane, with the lattice vectors becoming complex vectors in the Argand plane.
The modular symmetry in the upper half of the complex plane, $PSL(2,\mathbb{Z})$, has particularly nice features which rely on holomorphicity, the lack of complex conjugation symmetry, reminiscent of supersymmetry.

At first sight, modular symmetry does not look like a promising starting point for a family symmetry, for one thing it is an infinite group, since there are an infinite number of $2\times 2 $ matrices with integer elements and unit determinant. Secondly, it is not immediately obvious what a torus has got to do with particle physics. With the advent of superstring theory and extra dimensions, this second question may at least find an answer, since orbifold compactifications of two extra dimensions are often done on a torus~\cite{Ferrara:1989bc,Ferrara:1989qb}, and in superstring theory, the single lattice vector which describes the torus (in the convention that the other lattice vector has unit length and lies along the real axis) is promoted to the status of a field, called the modulus field $\tau$, where its vacuum expectation value (VEV) fixes the geometry of the torus \cite{Ishiguro:2021ccl,Cremades:2004wa,Ishiguro:2020tmo}. Moreover, it is possible to obtain a finite discrete group from the infinite modular group as discussed below.

The infinite modular group has a series of infinite normal subgroups called the principle congruence subgroups $\Gamma (N)$ of level $N$, whose elements are equal to the $2\times 2 $ unit matrix mod $N$ (where typically $N$ is an integer called the level of the group). 
For a given choice of level $N>2$, the quotient group $\Gamma_N=PSL(2,\mathbb{Z})/\Gamma (N)$ is finite and may be identified with the groups $\Gamma_N=A_4,S_4,A_5$ for levels $N=3,4,5$, which may be subsequently be used as a family symmetry~\cite{Feruglio:2017spp}. Indeed the only flavon present in such theories is the single modulus field $\tau$, whose VEV fixes the value of Yukawa couplings which form representations of $\Gamma_N$ and
are modular forms, leading to very predictive theories independent of flavons~\cite{Feruglio:2017spp}. 

Following the above observations~\cite{Feruglio:2017spp}, there has been considerable activity in applying modular symmetry to flavour models, and also in extending the framework to more general settings, following the bottom-up approach (see~\cite{Feruglio:2019ktm} for more details and extensive references). For example the modular $S_4$ group was studied in~\cite{Penedo:2018nmg, Novichkov:2020eep, Liu:2020akv}. To enhance the predictivity of such models, rather than considering the VEV of $\tau$ to be
a free complex parameter, it is interesting to consider fixed points or stabilizers
which are special values for the modulus field $\tau$ such as $\tau=i, \omega, i \infty$
where part of the modular transformations are preserved.
However such an approach with one modulus \footnote{Recently it has been claimed that a single modulus at $\tau=i$ can provide a good phenomenological description
of leptons, but this requires that the neutrino mass matrix is infinite at the fixed point~\cite{Feruglio:2022kea}.}
is rather too restrictive and generally calls for additional moduli fields 
which can be introduced 
in a straightforward way by considering additional modular groups, with one modulus per modular group,
as suggested in~\cite{deMedeirosVarzielas:2019cyj,King:2019vhv, deMedeirosVarzielas:2021pug, deMedeirosVarzielas:2022ihu,Devi:2023vpe}.
A recent example of a model of this kind was based on three finite modular groups 
$S_4^3$ broken to its diagonal subgroup $S_4$, with three moduli fields in the low energy theory 
located at three different fixed points, for example $\tau_1=i,\ \tau_2=i+2,\ \tau_3=\omega$,
leading to a very predictive and successful phenomenological description of the neutrino and charged lepton masses and lepton mixing
based on a version of the littlest seesaw~\cite{deMedeirosVarzielas:2022fbw}.

While there has been considerable effort devoted to studying modular symmetry arising from orbifolds in top-down heterotic string constructions~\cite{Baur:2019kwi}, \footnote{Top-down approaches suggest that the finite modular symmetry will typically be accompanied by a flavour symmetry leading to so called eclectic symmetry~\cite{Nilles:2020nnc,Nilles:2020kgo,Nilles:2020tdp,Baur:2020jwc,Ding:2023ynd}.}
there has been little work on bottom-up approaches which combine orbifolds together with modular symmetry.
In the bottom-up approach to modular symmetry as applied to flavour models, orbifolds are usually not considered at all.
Instead the formalism of modular symmetry and modular forms is adopted and flavour models then constructed,
without any reference to the underlying orbifold~\cite{Feruglio:2017spp}.
However there have been some bottom-up attempts to relate modular symmetry to orbifold GUTs, such as the model based on 
supersymmetric $SU(5)$ in 6d, where the two extra dimensions are compactified on a $\mathbb{T}^2/\mathbb{Z}_2$,
leading to a remnant $A_4$ with single modulus field located at the fixed point $ \tau=\omega$ of the orbifold~\cite{deAnda:2018ecu}.
In this model, there was also an $A_4$ flavour symmetry commuting with the $A_4$ modular symmetry, which was a pre-curser to 
the eclectic flavour symmetry approach~\cite{deAnda:2018ecu}.

In this paper we develop a bottom-up approach to flavour models which combines modular symmetry with orbifold constructions.
We shall consider orbifolds in 10d which can provide three modular groups and three moduli fields in the low energy theory (below the compactification scales). We assume that the 6 extra dimensions are factorisable into 3 tori, each defined by one complex coordinate $z_i$. Assuming supersymmetry, consistent with the holomorphicity requirement, 
we consider all the orbifolds of the form $(\mathbb{T}^2)^3/(\mathbb{Z}_N\times\mathbb{Z}_M)$,
and list all the available orbifolds, which have fixed values of the moduli fields (up to an integer).
The key advantage of such 10d orbifold models over 4d models is that the values of the moduli are not completely free but are 
constrained geometry and symmetry.

To illustrate the approach, we focus on the orbifold example $(\mathbb{T}^2)^3/(\mathbb{Z}_4\times\mathbb{Z}_2)$,
and discuss in detail the fixed points, with the choices $\tau_1=i,\ \tau_2=i+2$ being 
constrained by the orbifold, while $\tau_3$ is unconstrained but may be fixed by specifying a remnant symmetry.
Motivated by model building considerations we consider $\tau_3=\omega$ determined by imposing a remnant $S_4$ flavour symmetry.
We assume an $S_4^3$ modular symmetry, associated with each of the three moduli.
We show that such a model can reproduce a minimal 4d modular seesaw model of leptons
based on three finite modular groups 
$S_4^3$ broken to the diagonal modular subgroup $S_4$.
In the 4d models the three moduli fields
were simply assumed to lie at the fixed points, $\tau_1=i,\ \tau_2=i+2,\ \tau_3=\omega$~\cite{deMedeirosVarzielas:2022fbw},
but in the 10d model, these values are constrained by geometry and symmetry.
The resulting model is in the class
CSD$(n)$ with $n=1-\sqrt{6}$, where the atmospheric angle $\theta_{23}$ is restricted to the second octant.

The bottom-up approach to modular symmetry from orbifolds followed here 
can readily be extended to Grand Unified Theories (GUTs), with up to three moduli groups and moduli fields, including a remnant flavour symmetry, leading to a bottom-up version of the ecletic flavour symmetry in orbifold GUTs as anticipated in~\cite{deAnda:2018ecu}.

The layout of the remainder of the paper is as follows: In Sec. \ref{sec:orb}, the general SUSY preserving orbifolding is presented and shown how it fixes the modulus. This is shown for the case of 6 and 10 spacetime dimensions, as well as an specific detailed example. In Sec. \ref{sec:modsym} we describe the basics of the modular symmetry $S_4$, its corresponding modular forms at the fixed points as well as how it can arise as a remnant symmetry in orbifolding. In Sec. \ref{sec:model} we present a viable and predictive lepton model which uses $S_4^3$ modular symmetry in a $(\mathbb{T}^2)^3/(\mathbb{Z}_4\times\mathbb{Z}_2)$ orbifold. Finally in Sec. \ref{sec:conclusions} we present our conclusions.

\section{Orbifolding}
\label{sec:orb}

Modular symmetries have proved themselves very useful in model building. They may provide predictive flavor structure specially for the lepton sector without requiring the addition of extra fields nor complicated symmetry breaking mechanisms. A model with modular symmetry requires to be built in 6 dimensions (at least) and start with $\mathcal{N}=1$ SUSY, as the modular transformations are essentially transformations of the extra dimensional part of the enhanced Poincar\'e symmetry coupled with a SUSY transformation on the fields.

Most models assume a 6 dimensional spacetime with SUSY where the extra dimensions are compactified as a torus (with twist angle $\tau$) and build a model using the assumed modular symmetries. However assuming the extra dimensions to be a torus can't lead to a viable theory as the resulting model after compactification would have no chirality and $\mathcal{N}=2$ SUSY. The standard solution is to compactify the extra dimensions as an orbifold, which we now present its basics.

\subsection{The orbifold $\mathbb{T}^2/\mathbb{Z}_N$}

The two extra dimensional coordinates can be treated as a single complex coordinate $z=x_5+ix_6$. The torus compactification is done by identifying
\begin{equation}
z\sim z+1,\ \ \ z\sim z+\tau,
\label{eq:t2}
\end{equation}
where $\tau$ is called the twist angle and, for now, it is an arbitrary complex number. This identification restricts the range of the complex coordinate. The $\{1,\tau\}$ are called the basis vectors which generate the lattice of the extra dimensional plane and define the torus.

The torus by itself leads to a non chiral theory after compactification. The solution is to assume orbifolding, which is equivalent to assume that the extra dimensional part of the Poincar\'e group is not a full symmetry. This is done by modding out a discrete subgroup of the extra dimensional Lorentz group, which is called orbifolding. In 6 dimensions, the extra dimensional part of the Lorentz group is 
\begin{equation}
SO(1,5)/SO(1,3)\simeq SO(2)\simeq U(1),
\end{equation}
which correspond to rotation in the 2 extra dimensions. One can mod out by any discrete subgroup $F\in U(1)$, which can only be $F=\mathbb{Z}_N$, with $N$ an arbitrary integer (for now). It has to be a discrete group to avoid reducing the dimensionality. The $\mathbb{Z}_N$ orbifolding is achieved by the identification
\begin{equation}
z\sim e^{2i\pi/N} z,
\label{eq:t2zn}
\end{equation}
which further restricts the range of the extra dimensional coordinates. The orbifold has fixed points which allow boundary conditions that generate chirality, may break the gauge symmetry and reduce the enhanced SUSY. Therefore they may lead to a consistent model after compactification.

To avoid dimensional reduction and therefore for the orbifold to be consistent, the orbifold action in Eq.\ref{eq:t2zn} must be equivalent to an integer number of lattice transformations as in Eq. \ref{eq:t2zn}. In other words, there must exist integer numbers $a,b\in \mathbb{Z}$ such that a solution exists for
\begin{equation}
e^{2i\pi/N} z=z+a+b\tau.
\end{equation}
It is enough to find a solution for each of the basis vectors $\{1,\tau\}$,
\begin{equation}
e^{2i\pi/N}=a+b\tau,\ \ \ e^{2i\pi/N}\tau =c+d\tau,
\end{equation}
where there must exist $a,b,c,d \in \mathbb{Z}$ that solve these equations.
It is clear that there is no solution for arbitrary $N$ and $\tau$. This restricts the $N$ and $\tau$ to be one of
\begin{equation}
\begin{split}
N=2,&\ \ \ \tau = z\in \mathbb{C},\\
N=3,&\ \ \ \tau = \omega,\\
N=4,&\ \ \ \tau = i,\\
N=6,&\ \ \ \tau = \{\omega,\rho/\sqrt{3}\},\\
\end{split}
\end{equation}
where $\omega  = e^{2i\pi/3}$ and $\rho  = e^{i\pi/6}$ and all the solutions are valid up to an integer.

Therefore working with an orbifold may fix $\tau$ geometrically, adding predictivity, and solves the chirality problem therefore allowing a viable model.

\subsection{The orbifold $(\mathbb{T}^2)^3/(\mathbb{Z}_N\times\mathbb{Z}_M)$}
\label{sec:orbs}

Many models may require various independent modular symmetries or different $\tau$ values to achieve a better fit. One such model is presented in Sec. \ref{sec:model}.
As it needs 3 independent modular symmetries, we focus on 10 dimensional spaces with $\mathcal{N}=1$ SUSY before and after compactification.

In the 10 dimensional case, one can orbifold by a discrete subgroup of the extra dimensional part of the Lorentz group
\begin{equation}
SO(1,9)/SO(1,3)\simeq SO(6)\simeq SU(4),
\end{equation}
which corresponds to rotations in the extra 6 dimensions. The former $SU(4)$ can be identified with the $SU(4)_\mathcal{R}$ of the enhanced $\mathcal{N}=4$ SUSY. As we want to preserve simple SUSY after compactification, the discrete orbifolding group must be $F\subset SU(3)$. As it is rank 2,
a general 10d SUSY preserving abelian factorisable orbifolding is
\begin{equation}
(\mathbb{T}^2)^3/(\mathbb{Z}_N\times\mathbb{Z}_M)
\end{equation}
which can be compactified by the basis vectors
\begin{equation}
z_i\sim z_i+1,\ \ \ z_i\sim z_i+\tau_i,
\label{eq:traorb}
\end{equation}
and the orbifolding defined by
\begin{equation}
\begin{split}
\theta_N:\ (x,z_1,z_2,z_3)&\sim  (x,\alpha_N z_1,\beta_N z_2,\gamma_N z_3),\\
\theta_M:\ (x,z_1,z_2,z_3)&\sim  (x,\alpha_M z_1,\beta_M z_2,\gamma_M z_3),
\label{eq:rotorb}
\end{split}
\end{equation}
where $\alpha_{N,M},\beta_{N,M},\gamma_{N,M}$ are Nth, Mth roots of unity.

The choice of the phases of the orbifolding are restricted by the preservation of $\mathcal{N}=1$ SUSY. The $\tau_i$  must be fixed so that the lattice is unchanged by the orbifold transformation.
The $\tau_i$ are fixed, as they must such that the orbifolding identification does not change the lattice and therefore the torus remains unchanged. Therefore there must exist integers $a,b,c,d$ such that
\begin{equation}
\begin{split}
(\delta,\delta\tau)&=(a+b\tau,c+d\tau),
\label{eq:orbres}
\end{split}
\end{equation}
for each corresponding $\delta=\alpha_{N,M},\beta_{N,M},\gamma_{N,M}$

These restrictions limit the available (SUSY preserving \cite{Fischer:2012qj}) orbifolds to be as in Table \ref{tab:orb},
which displays all the available orbifolds with some of the $\tau_i$ fixed as shown (up to an integer), while the non-fixed
values are indicated by the complex number $z$.
\begin{table}[h!]
\begin{center}
\begin{tabular}{l|lll}
$(N,M)$ & $(\alpha_N,\beta_N,\gamma_N)$ &$ (\alpha_M,\beta_M,\gamma_M )$& $(\tau_1,\tau_2,\tau_3)$ \\
\hline\\
$(3,1)$& $(\omega,\omega,\omega)$ & $(1,1,1)$ &$ (\omega,\omega,\omega)$\\
$(4,1)$& $(i,i,-1)$ &$ (1,1,1)$ &$ (i,i,z)$ \\
$(6,1)_I$&$ (-\omega^2,-\omega^2,\omega^2)$ &$ (1,1,1)$ & $(\{\omega,\rho/\sqrt{3}\},\{\omega,\rho/\sqrt{3}\},\omega)$\\
$(6,1)_{II}$&$ (-\omega^2,\omega,-1)$ & $(1,1,1)$ & $(\{\omega,\rho/\sqrt{3}\},\omega,z)$\\
$(2,2)$& $(1,-1,-1)$ & $(-1,1,-1)$ & $(z,z,z)$\\
$(4,2)$& $ (i,-i,1)$ &$(1,-1,-1)$ & $(i,i,z)$\\
$(6,2)_I$& $(-\omega^2,1,-\omega) $& $(1,-1,-1) $& $(\{\omega,\rho/\sqrt{3}\},z,\{\omega,\rho/\sqrt{3}\})$\\
$(6,2)_{II}$& $(\omega^2,-\omega^2,-\omega^2)$ &$ (1,-1,-1)$ & $(\omega,\{\omega,\rho/\sqrt{3}\},\{\omega,\rho/\sqrt{3}\})$\\
$(3,3)$& $(1,\omega,\omega^2)$ &$ (\omega,1,\omega^2)$ & $(\omega,\omega,\omega)$\\
$(6,3)$& $(-\omega^2,1,-\omega) $&$ (1,\omega,\omega^2)$ & $(\{\omega,\rho/\sqrt{3}\},\{\omega,\rho/\sqrt{3}\},\omega)$\\
$(4,4)$& $(1,i,-i)$ & $(i,1,-i) $& $(i,i,i)$\\
$(6,6)$& $(1,-\omega^2,-\omega)$ & $(-\omega^2,1,-\omega)$ & $(\{\omega,\rho/\sqrt{3}\},\{\omega,\rho/\sqrt{3}\},\{\omega,\rho/\sqrt{3}\})$\\
\end{tabular}
\caption{\label{tab:orb} Comprehensive list of 6d abelian factorisable and SUSY preserving orbifolds $(\mathbb{T}^2)^3/(\mathbb{Z}_N\times\mathbb{Z}_M)$, where $\omega  = e^{2i\pi/3}$ and $\rho  = e^{i\pi/6}$, and the fixed points of $\tau_i$ are specified only up to an integer. For example, $\tau_2=i,i+1,i+2$ and so on are all equivalent.
The values of the complex numbers $z$ are not restricted by the orbifold, but particular values of $z$ may be fixed by a remnant global symmetry.}
\end{center}
\end{table}

\subsection{The orbifold $(\mathbb{T}^2)^3/(\mathbb{Z}_4\times\mathbb{Z}_2)$ }
\label{sec:z4z2}
In this subsection we discuss an example of an orbifold chosen from 
Table \ref{tab:orb} corresponding to $(N,M)=(4,2)$ which leads to an interesting model
\footnote{This example is not unique, there are other choices which also lead to viable models.}.
The full model based on the resulting orbifold $(\mathbb{T}^2)^3/(\mathbb{Z}_4\times\mathbb{Z}_2)$
will be presented in Sec. \ref{sec:model}.
The model we have in mind is an extra dimensional version of a four dimensional model based on three finite modular groups 
$S_4^3$ broken to a diagonal subgroup $S_4$, with the three moduli fields in the low energy theory 
located at three different fixed points, namely $\tau_1=i,\ \tau_2=i+2,\ \tau_3=\omega$.
In a 4d framework, this was shown to lead to a very predictive and successful phenomenological description of the neutrino and charged lepton masses and lepton mixing
based on a type of littlest seesaw~\cite{deMedeirosVarzielas:2022fbw}.

In the 10d framework considered here, the desired moduli fields $\tau_i$ for such model are in principle consistent with the orbifold divisors
$\mathbb{Z}_2\times \mathbb{Z}_2,\ \mathbb{Z}_4,\ \mathbb{Z}_4\times \mathbb{Z}_2$. However 
$\mathbb{Z}_2\times \mathbb{Z}_2$ does not fix any of the $\tau_i$, so is not so restrictive. The $\mathbb{Z}_4$ orbifold divisor fixes the $\tau_i$ as needed by the model, but does not have the necessary fixed branes to build consistent interactions. We are then left with the only viable and predictive choice being the orbifold divisor $\mathbb{Z}_4\times \mathbb{Z}_2$, which can lead to the desired fixed points, as we discuss below.

We assume, then, a 10d spacetime where the 6 extra dimensions are factorisable into 3 tori, each defined by one complex coordinate $z_i$ with $i=1,2,3,$ and compactified as in Eq. \ref{eq:traorb} 
\begin{equation}
z_i\sim z_i+1,\ \ \ z_i\sim z_i+\tau_i,
\end{equation}
The orbifold $(\mathbb{T}^2)^3/\mathbb{Z}_4\times\mathbb{Z}_2$
as defined by the orbifolding actions in Eq. \ref{eq:rotorb}, using Table~\ref{tab:orb} with $(N,M)=(4,2)$ then implies,
\begin{equation}
\begin{split}
\theta_4&: (x,z_1,z_2,z_3)\sim (x,iz_1,-iz_2,z_3),\\
\theta_2&: (x,z_1,z_2,z_3)\sim (x,z_1,-z_2,-z_3).
\label{eq:orbac}
\end{split}
\end{equation}

In the orbifold approach, $(1,\tau_i)$ define the twist and the basis vectors of each torus.
For the orbifold to be consistent, the orbifolding actions $\theta_{2,4}$ must not change the lattice, i.e. its action over the lattice basis vectors $(1,\tau_i)$ must be a linear combination of the original lattice vectors, with integer coefficients.  
 Therefore there must exist integers $a_{1,2,3},b_{1,2,3},c_{1,2,3},d_{1,2,3}\in \mathbb{Z}$ such that,
as in Eq. \ref{eq:orbres}
\begin{equation}
\begin{split}
(i,i\tau_{1,2})&=(a_{1,2}+b_{1,2}\tau_{1,2},c_{1,2}+d\tau_{1,2}),\\
(-1,-\tau_{3})&=(a_{3}+b_{3}\tau_{3},c_{3}+d\tau_{3}),\\
\end{split}
\label{1}
\end{equation}
In the present example, solving Eq.~\ref{1} gives,
\begin{equation}
\begin{split}
\tau_{1,2} &=i+n_{1,2},\ \ \ | \ \ \ n_{1,2}\in \mathbb{Z},\\
\tau_3 &\in \mathbb{C}.
\label{eq:tau24}
\end{split}
\end{equation}
which corresponds to the result given in Table~\ref{tab:orb} with $(N,M)=(4,2)$.
We emphasise that the twists $\tau_i$ are fixed geometrically by the orbifold actions. Therefore in the orbifold approach to modular symmetries,
the moduli fields are not a completely free choice, but are constrained as in Table \ref{tab:orb}.

Each orbifold action in Eq. \ref{eq:orbac}, leaves some invariant subspaces which are called fixed branes
\begin{equation}
\begin{split}
\theta_4&:\left(x,\left\{0,\frac{i+1}{2}\right\},\left\{0,\frac{i+1}{2}\right\},z_3\right),\\
\theta^2_4 &:\left(x,\left\{0,\frac{1}{2},\frac{i}{2},\frac{i+1}{2}\right\},\left\{0,\frac{1}{2},\frac{i}{2},\frac{i+1}{2}\right\},z_3\right),\\
\theta_2 &:\left(x,z_1,\left\{0,\frac{1}{2},\frac{i}{2},\frac{i+1}{2}\right\},\left\{0,\frac{1}{2},\frac{\tau_3}{2},\frac{\tau_3+1}{2}\right\}\right),\\
\theta_2 \theta_4 &:\left(x,\left\{0,\frac{i+1}{2}\right\},\left\{0,\frac{i+1}{2}\right\},\left\{0,\frac{1}{2},\frac{\tau_3}{2},\frac{\tau_3+1}{2}\right\}\right),\\
\theta_2 \theta_4^2&:\left(x,\left\{0,\frac{1}{2},\frac{i}{2},\frac{i+1}{2}\right\},z_2,\left\{0,\frac{1}{2},\frac{\tau_3}{2},\frac{\tau_3+1}{2}\right\}\right).
\label{eq:branes}
\end{split}
\end{equation}
When building a model, fields can be chosen to be located in any of the previous branes or in the bulk.

We want a minimal model where all fields can behave as modular forms (with different $\tau_i$ depending on their location) but can interact with each other, we will only use the 6d branes
\begin{equation}
\begin{split}
\mathbb{T}^2_A &=(x,z_1,0,0),\\
\mathbb{T}^2_B &=(x,0,z_2,0),\\
\mathbb{T}^2_C &=(x,0,0,z_3),
\end{split}
\end{equation}
where all of them touch at the origin brane, where all interactions happen.

From Eq. \ref{eq:orbac}, we note that the $z_1$ only feels the $\theta_4$ action, therefore the $\mathbb{T}^2_A $ is a $\mathbb{Z}_4$ orbifold. As the action of $\theta_2$ on $z_2$ is also contained in $\theta_4$, the $\mathbb{T}^2_B $ is also a $\mathbb{Z}_4$ orbifold. Finally the $z_3$ only feels the $\theta_2$ action, therefore the $\mathbb{T}^2_C $ is a $\mathbb{Z}_2$ orbifold.

\section{Modular $S_4$ symmetries in the orbifold approach}
\label{sec:modsym}

So far we have considered possible orbifolds in which the VEVs of the moduli fields $\tau_i$ are fixed at least partially by the geometry.
We now turn to the modular symmetries of the fields $\tau_i$ which are broken by the VEVs of the moduli fields $\tau_i$.
In general such modular symmetries are infinite but have a series of infinite normal subgroups called the principle congruence subgroups 
$\Gamma (N)$ of level $N$, whose elements are equal to the $2\times 2 $ unit matrix mod $N$ (where typically $N$ is an integer called the level of the group). 

These matrix modular transformations are applied to the 2 extra dimensional basis vectors $\{1,\tau\}$ and are such that the lattice these vectors generate remains invariant. In this work we study 10 dimensional orbifolds where we restrict ourselves to the case where the 6 extra dimensions are factorisable into 3 independent tori $\mathbb{T}^2_1\times\mathbb{T}^2_2\times\mathbb{T}^2_3$. Each torus generated by it own set of basis vectors $\{1,\tau_i\}$ and therefore each of them has an independent modular symmetry, making the general modular symmetry the direct product of each one corresponding to each torus.

For a given choice of level $N>2$, the quotient group $\Gamma_N=PSL(2,\mathbb{Z})/\Gamma (N)$ is finite and may be identified with the groups $\Gamma_N=A_4,S_4,A_5$ for levels $N=3,4,5$, which may be subsequently be used as a family symmetry~\cite{Feruglio:2017spp}. In this section we consider the case $N=4$ which corresponds to modular $S_4$ symmetries.

With two extra dimensions the single complex modulus $\tau$ has an infinite modular symmetry $\overline{\Gamma}=SL(2,\mathbb{Z})$ as follows.
The modular group $\overline{\Gamma}$ is the group of linear fraction transformations which acts on the complex modulus $\tau$ in the upper half complex plane as follow,
\begin{equation}
\tau\rightarrow\gamma\tau=\frac{a\tau+b}{c\tau+d},~~\text{with}~~a, b, c, d\in\mathbb{Z},~~ad-bc=1,~~~\Im\tau>0\,.
\end{equation}
The modular group $\overline{\Gamma}$ can be generated by two generators $S$ and $T$
\begin{equation}
S:\tau\mapsto -\frac{1}{\tau},~~~~\quad T: \tau\mapsto\tau+1.
\end{equation}

From the infinite modular group the finite subgroup $\Gamma_N=PSL(2,\mathbb{Z})/\Gamma (N)$ may be obtained.
A crucial element of the modular invariance approach is the modular form $f(\tau)$ of weight $k$ and level $N$. The modular form $f(\tau)$ is a holomorphic function of the complex modulus $\tau$ and it is required to transform under the action of $\overline{\Gamma}(N)$ as follows,
\begin{equation}
f\left(\frac{a\tau+b}{c\tau+d}\right)=(c\tau+d)^kf(\tau)
\end{equation}
The modular forms of level $N=4$ have been constructed in~\cite{Penedo:2018nmg,Novichkov:2018ovf} .

The associated finite modular group $\Gamma_4$ has two generators $S$ and $T$ which fulfill the following  rations
\begin{equation}
S^2=(ST)^3=(TS)^3=T^4=1\,.
\end{equation}
The finite modular group $\Gamma_4$ is isomorphic to the permutation group $S_4$ of four objects. In order to see the correlation between $S_4$ and tri-bimaximal mixing and the connection to $S_3$, $A_4$ groups more easily, it is convenient to generate the $S_4$ group in terms of three generators $\hat{S}$, $\hat{T}$ and $\hat{U}$ with the multiplication rules~\cite{Hagedorn:2010th,Ding:2013hpa},
\begin{eqnarray}
\hat{S}^2=\hat{T}^3=\hat{U}^2=(\hat{S}\hat{T})^3=(\hat{S}\hat{U})^2=(\hat{T}\hat{U})^2=(\hat{S}\hat{T}\hat{U})^4=1\,,
\label{eq:s4stu}
\end{eqnarray}
where $\hat{S}$ and $\hat{T}$ alone generate the group $A_4$, while $\hat{T}$ and $\hat{U}$ alone generate the group $S_3$. The generators $S$, $T$ can be expressed in terms of $\hat{S}$, $\hat{T}$ and $\hat{U}$ \begin{equation}
S = \hat{S}\hat{U},\, T = \hat{S} \hat{T}^2 \hat{U},\,ST = \hat{T}
\end{equation}
or vice versa
\begin{equation}
\hat{S}=(ST^2) ^2,~~~ \hat{T}=ST,~~~\hat{U}=T^2ST^2\,,
\end{equation}
with the explicit matrices being
\begin{equation}
\hat{S}=\frac{1}{3}\left(\begin{array}{ccc}
 -1 ~& 2  &~ 2  \\
 2  ~& -1  &2 \\
 2  ~& 2 &~ -1
\end{array}\right),\ \ \ 
\hat{T}= \left(\begin{array}{ccc}
 1 ~& 0  &~ 0  \\
 0  ~& \omega^2  &0 \\
 0  ~& 0 &~ \omega
\end{array}\right),\ \ \ 
\hat{U}= \mp \left(\begin{array}{ccc}
 1 ~& 0  &~ 0  \\
 0  ~& 0 &1 \\
 0  ~& 1 &~ 0
\end{array}\right),
\end{equation}
where the minus sign in $\hat{U}$ applies for the $\textbf{3}$ representation while the plus sign is for the $\textbf{3}'$ representation.

\begin{table}[h]
\begin{center}
\begin{tabular}{|c|c|c|}\hline\hline
 ~~  &  $S$  &   $T$     \\ \hline
~~~$\mathbf{1}$, $\mathbf{1}^\prime$ ~~~ & $\pm1$   &  $\pm1$  \\ \hline
   &   &       \\ [-0.16in]
$\mathbf{2}$ &  $\left( \begin{array}{cc}
 0&~1 \\
 1&~0
 \end{array} \right) $
 & $\left( \begin{array}{cc}
 0 ~&~ \omega^2 \\
 \omega ~&~ 0
\end{array} \right) $
    \\ [0.15in]\hline
   &   &       \\ [-0.16in]
$\mathbf{3}$, $\mathbf{3}^\prime $ & $\pm\frac{1}{3} \left(\begin{array}{ccc}
 1 ~& -2  &~ -2  \\
 -2  ~& -2  &~ 1 \\
 -2  ~& 1 &~ -2
\end{array}\right)$ &
$\pm\frac{1}{3}\left( \begin{array}{ccc}
1 ~& -2\omega^2 &~ -2\omega \\
-2 ~& -2\omega^2 &~ \omega \\
-2  ~& \omega^2 &~ -2\omega
\end{array}\right) $
\\[0.25in] \hline\hline
\end{tabular}
\caption{\label{tab:S4_rep_hat}The representation matrices of the generators $S$ and $T$ in the five irreducible representations of $S_4$, where $\omega=e^{2\pi i/3}=-1/2+i\sqrt{3}/2$ is a cubic root of unity. }
\end{center}
\end{table}

We assume $\mathcal{N}=1$ SUSY in 10d and this abelian orbifold preserves $\mathcal{N}=1$ SUSY in 4d after compactification \cite{Fischer:2012qj}.
Therefore we can assume 3 independent modular symmetry groups, each associated with a different 
tori~\cite{deMedeirosVarzielas:2019cyj,King:2019vhv, deMedeirosVarzielas:2021pug, deMedeirosVarzielas:2022ihu}.
We assume three discrete modular symmetries $S_4^{A,B,C}$ associated to each complex coordinate $z_{1,2,3}$ correspondingly.

With the assumed $S_4$ modular symmetries, the corresponding moduli from Eq.\ref{eq:tau24}, which have an arbitrary integer, now can only be 
\begin{equation}
n=0,1,2,3,
\end{equation}
where it is now limited to a choice of one in four.

\subsection{Fixed points and $S_4$ modular forms}

In most models using modular symmetries, the $\tau$ is a free parameter that is minimized by a potential and treated as a VEV. A standard strategy to increase the predictivity of the model is to restrict to fixed points which are geometrically preferred. These point $\bar{\tau}$ are defined as the points that are invariant under some element of the modular group $\gamma\in S_4$ called the stabilizer.

In an orbifold, the $\tau$ is not a free parameter and it is fixed by the geometry of the orbifold itself. However, there are a finite number of choices, which allow specific modular forms which are listed in Table \ref{ta:modfor}  \cite{Ding:2019gof}. All the presented $S_4$ modular forms are defined in the basis from Table \ref{tab:S4_rep_hat}.

\begin{table}[h]
\begin{center}
\resizebox{1.0\textwidth}{!}{
\begin{tabular}{|c|c|c|c|c|}\hline
 $\tau$ & \multicolumn{2}{c|}{$Y^{(2)}_\mathbf{3}(\tau)$, $Y^{(6)}_\mathbf{3,I}(\tau)$} & $Y^{(4)}_\mathbf{3}(\tau)$, $Y^{(6)}_\mathbf{3'}(\tau)$ & $Y^{(4)}_\mathbf{3'}(\tau)$, $Y^{(6)}_\mathbf{3,II}(\tau)$ \\
\hline\hline
 $i$ & \multicolumn{2}{c|}{$(1,1+\sqrt{6},1-\sqrt{6})$} & $(1,-\frac{1}{2},-\frac{1}{2})$ & $(1,1-\sqrt{\frac{3}{2}},1+\sqrt{\frac{3}{2}})$  \\
\hline
 $i+1$ & \multicolumn{2}{c|}{$(1,-\frac{\omega}{3}(1+i\sqrt{2}),-\frac{\omega^2}{3}(1+i\sqrt{2}))$} & $(0,1,-\omega)$ & $(1,\frac{i\omega}{\sqrt{2}},\frac{i\omega^2}{\sqrt{2}})$  \\
\hline
$i+2$ & \multicolumn{2}{c|}{$(1,\frac{1}{3}(-1+i\sqrt{2}),\frac{1}{3}(-1+i\sqrt{2}))$} & $(0,1,-1)$ & $(1,-\frac{i}{\sqrt{2}},-\frac{i}{\sqrt{2}})$  \\
\hline
 $i+3$ & \multicolumn{2}{c|}{$(1,\omega(1+\sqrt{6}),\omega(1-\sqrt{6}))$} & $(1,-\frac{\omega}{2},-\frac{\omega^2}{2})$ & $(1,\omega(1-\sqrt{\frac{3}{2})},\omega^2(1+\sqrt{\frac{3}{2}}))$ \\
\hline\hline
$\tau$ & $Y^{(2)}_\mathbf{3}(\tau)$ & $Y^{(4)}_\mathbf{3}(\tau)$,$Y^{(4)}_\mathbf{3'}(\tau)$ & $Y^{(6)}_\mathbf{3,II}(\tau)$,$Y^{(6)}_\mathbf{3'}(\tau)$ & $Y^{(6)}_\mathbf{3,I}(\tau)$
\\ \hline \hline
 $\omega$ & $(0,1,0)$ & $(0,0,1)$ & $(1,0,0)$ & \multirow{7}*{$(0,0,0)$}\\
\cline{1-4}
 $\omega+1$ & $(1,1,-\frac{1}{2})$ & $(1,-\frac{1}{2},1)$& $(1,-2,-2)$  &  \\
\cline{1-4}
 $\omega+2$ & $(1,-\frac{\omega^2}{2},\omega)$ & $(1,\omega^2,-\frac{\omega}{2})$ & $(1,-2\omega^2,-2\omega)$& \\
\cline{1-4}
 $\omega+3$ & $(1,\omega,-\frac{\omega^2}{2})$ & $(1,-\frac{\omega}{2},\omega^2)$ & $(1,-2\omega,-2\omega^2)$ & \\
\cline{1-4}
 $\rho/\sqrt{3}$ & $(1,-\frac{\omega}{2},\omega^{2})$ & $(1,\omega,-\frac{\omega^{2}}{2})$ & $(1,-2\omega,-2\omega^{2})$ & \\ 
 \cline{1-4}
  $\rho/\sqrt{3}+1$ & $(0,0,1)$ & $(0,1,0)$ & $(1,0,0)$ & \\ 
 \cline{1-4}
   $\rho/\sqrt{3}+2$ & $(1,-\frac{1}{2},1)$ & $(1,1,-\frac{1}{2})$ & $(1,-2,-2)$ & \\ 
 \cline{1-4}
 $\rho/\sqrt{3}+3$ & $(1,\omega^2,-\frac{\omega}{2})$ & $(1,-\frac{\omega^2}{2},\omega)$ &$(1,-2\omega^2,-2\omega)$ & \\
\hline
\end{tabular} }
\caption{\label{ta:modfor} The alignments of triplet modular forms $Y_{\mathbf{3}, \mathbf{3'}}(\tau)$ of level 4 up to weight 6 with the available fixed moduli in orbifolds. We have ignored the overall constant appearing in each alignment.}
\end{center}
\end{table}

In the $(\mathbb{T}^2)^3/(\mathbb{Z}_4\times\mathbb{Z}_2)$ orbifold, it will be assumed that 
\begin{equation}
\tau_1=i,\ \ \ \tau_2=i+2,
\end{equation}
which are particular cases of Eq.~\ref{eq:tau24} which 
are phenomenologicaly preferred, as described in the Sec. \ref{sec:model}. 
However the choice of $\tau_3$ is undetermined by the $(\mathbb{T}^2)^3/(\mathbb{Z}_4\times\mathbb{Z}_2)$ orbifold,
and instead shall be fixed by assuming a remnant $S_4$ symmetry, as discussed in the next subsection.

\subsection{$S_4$ Remnant Symmetry}
\label{sec:rems4}

The orbifold $\mathbb{Z}_2$, associated with the third torus $\mathbb{T}^2_C$, does not fix $\tau$. 
However, supposing that the twist angle is $\tau=\omega=e^{2i\pi/3}$ would leave a remnant $S_4$ symmetry (which is a subgroup of the extra dimensional Poincar\'e group) after compactification \cite{deAnda:2018oik,Altarelli:2006kg}. We shall assume that there is a remnant $S_4$ after compactification, therefore fixing uniquely
\begin{equation}
\tau_3=\omega.
\end{equation}

We focus on the branes of the fixus torus $\mathbb{T}_C$ \cite{Altarelli:2006kg,Adulpravitchai:2009id,Adulpravitchai:2010na},
\begin{equation}
\bar{z}= \{0,\ 1/2,\ \omega/2,\ \omega^2/2\},
\end{equation}
which are naturally invariant under the orbifold transformations
\begin{equation}
T_1:\ \bar{z}\to \bar{z}+1,\ \ \ T_2:\ \bar{z}\to \bar{z}+\omega,\ \ \ Z:\bar{z}\to -\bar{z}.
\end{equation}

The set of branes is invariant under the permutation set of them. However not all permutations are Poincar\'e transformations.

These fixed branes and are permuted by the Poincar\'e transformations
\begin{equation}
S_1:\bar{z}\to \bar{z}+1/2,\ \ \ S_2:\bar{z}+\omega/2,\ \ \ R:\bar{z}\to\omega \bar{z},\ \ \ P:\bar{z}\to \bar{z}^*,\ \ \ P':\bar{z}\to-\bar{z}^*,
\end{equation}
which, after orbifolding, generate the remnant symmetry. We can write these operations explicitly $S_1[(12)(34)],\ S_2[(13)(24)],\ R[(243)(1)],\ P[(34)(1)(2)],\ P'[(34)(1)(2)].$
There are only 3 independent transformations since $S_2=R^2\cdot S_1\cdot R,\ \ P=P'$. 

These symmetry transformations relate to the $S_4$ generators with $\hat{S}=S_1,\ \hat{T}=R,\ \hat{U}=P$ satisfying Eq. \ref{eq:s4stu}
which is the presentation rules for the $S_4$ symmetry \cite{King:2013eh}. 

The $\mathbb{T}_{A,B}$ have only 2 branes from Eq. \ref{eq:branes}. Therefore its remnant symmetry can only be $\mathbb{Z}_2$.

With the assumption of an $S_4$ remnant symmetry, the $\tau_3$ is fixed geometrically 
to be equal to $\omega$ \cite{deAnda:2018ecu}. .

\section{ A realistic orbifold model}
\label{sec:model}

\begin{figure}[h!]
	\centering
	\begin{subfigure}{0.45\textwidth}
		\centering
		\includegraphics[scale=0.45]{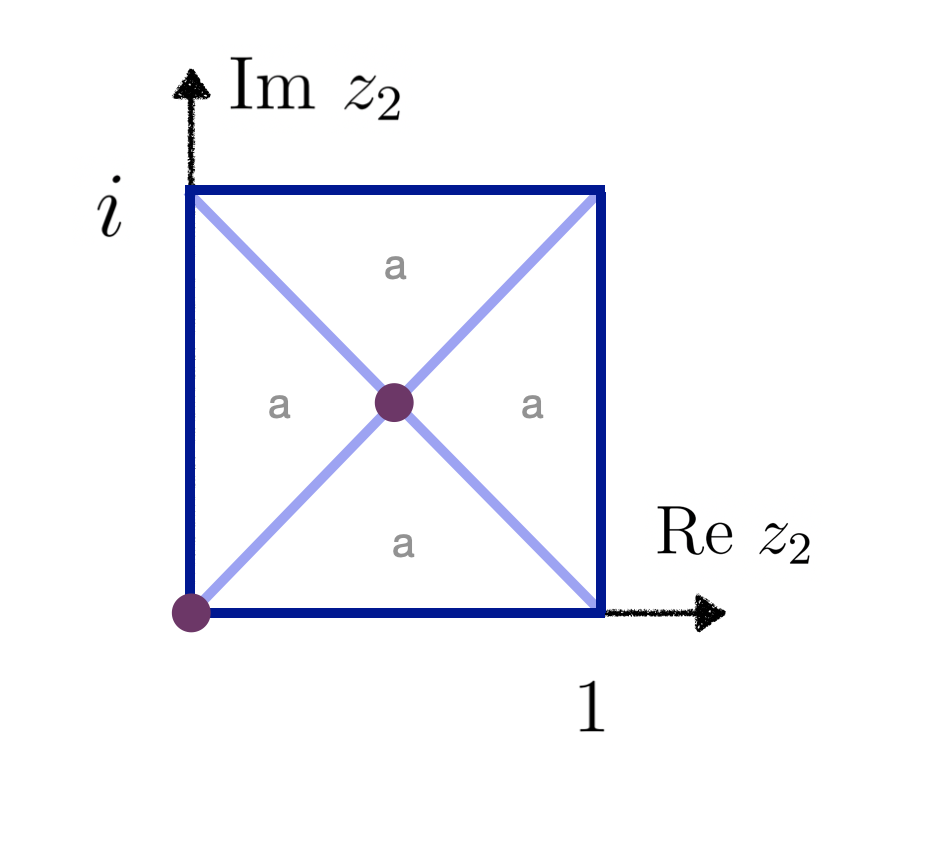}
		\caption*{(a) The extra dimensional space for $\mathbb{T}^2_A$. The $\mathbb{Z}_4$ orbifolding identifies the four isosceles triangles labeled as $a$.}
	\end{subfigure}%
	\ \ \ \ \
	\begin{subfigure}{0.45\textwidth}
		\centering
		\includegraphics[scale=0.45]{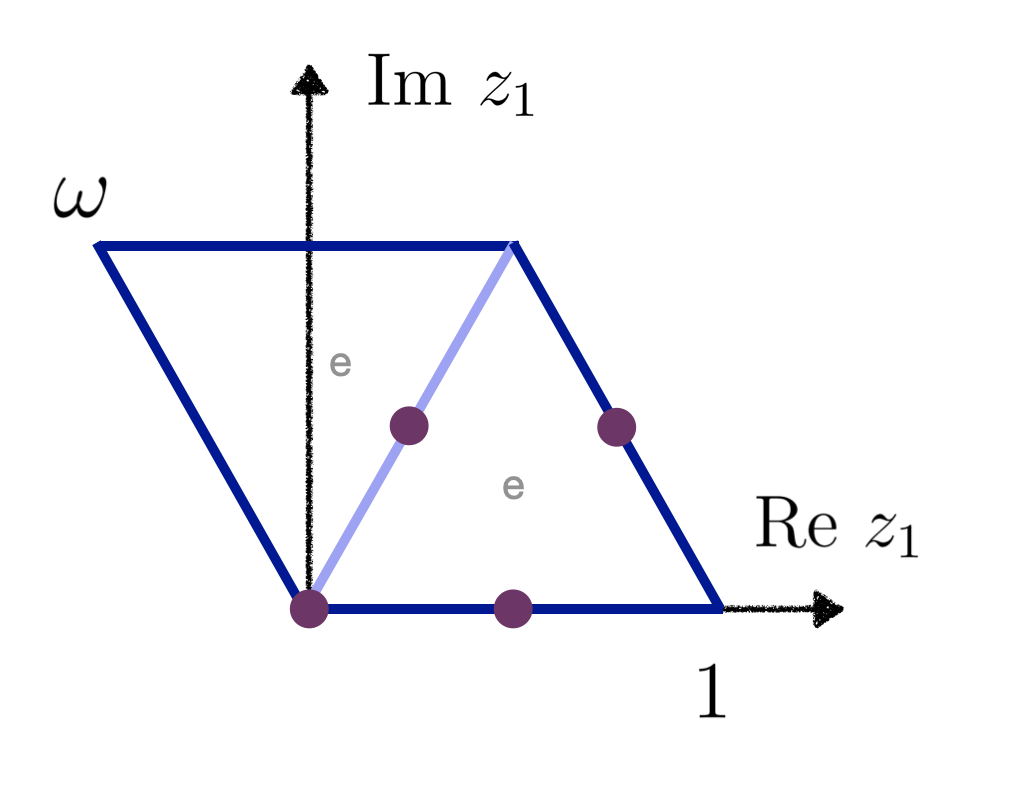}
		\caption*{(c) The extra dimensional space for $\mathbb{T}^2_C$. I The $\mathbb{Z}_2$ orbifolding identifies the two equilateral triangles labeled as $e$.}
	\end{subfigure}
	\begin{subfigure}{0.95\textwidth}
		\centering
		\includegraphics[scale=0.45]{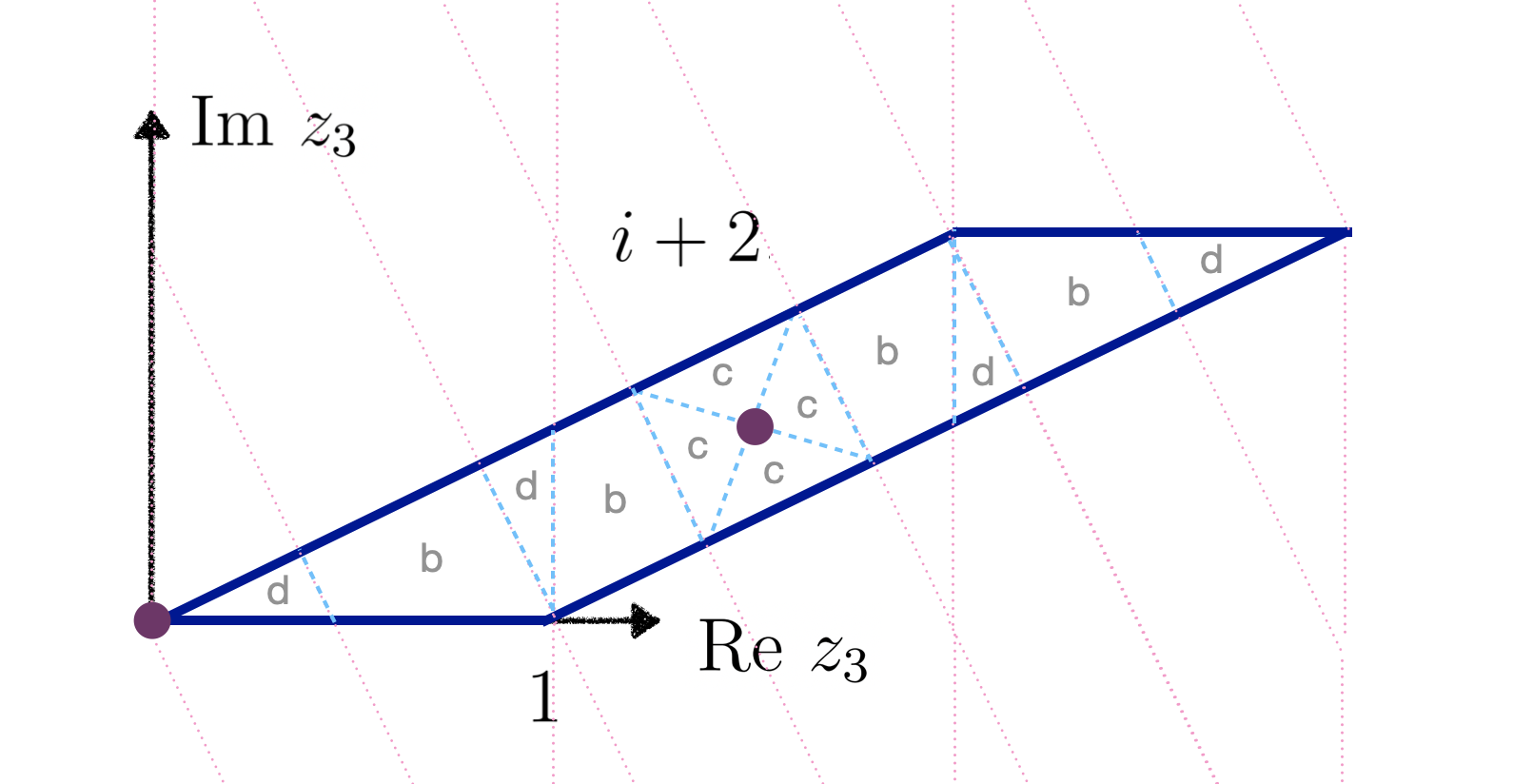}
		\caption*{(b) The extra dimensional space for $\mathbb{T}^2_B$.  The $\mathbb{Z}_4$ orbifolding is done by rotating the space by $\pi/2$ and creating drawing the lattice (dotted pink). One identifies the overlaps, which is this case are four quadrilaterlas labeled as $b$, four isosceles triangles labeled as $c$ and four right angle triangles labeled as $d$.}
	\end{subfigure}%
	\caption{Visualization of the extra dimensional space for each of the fundamental tori $\mathbb{T}^2_{A,B,C}.$ Identifying together opposite sides we obtain $\mathbb{T}^2$. The orbifolding is described in each subfigure. The dots represent the fixed points.}
	\label{fig:orbs}
\end{figure}

We now turn to a concrete 10d bottom-up orbifold model with three factorizable tori built from the fundamental space depicted geometrically in Fig. \ref{fig:orbs}.
The 10d model is compactified on an orbifold
$(\mathbb{T}^2)^3/(\mathbb{Z}_4\times\mathbb{Z}_2)$ and we assume
three finite modular symmetries $S^{A,B,C}_4$. Furthermore there is a remnant $S_4$ symmetry 
whose only role is to fix $\tau_3=\omega$ \footnote{As discussed later, remnant $S_4$ symmetry may be further employed to 
control the K\"ahler potential.}.
This uniquely fixes the moduli geometrically to be $\tau_1=i,\ \tau_2=i+2,\ \tau_3=\omega,$ (up to a choice in four).

The field content which defines the model is given in Table~\ref{ta:model}.

\begin{table}[h]
\centering
\begin{footnotesize}
 \begin{tabular}{| l | c c c c c c|c|}
\hline \hline
Field & $S_4^A$ & $S_4^B$ & $S_4^C$ & \!$2k_A$\! & \!$2k_B$\! & \!$2k_C$\! & Loc\\ 
\hline \hline
$L$ & $\mathbf{1}$ & $\mathbf{1}$ & $\mathbf{3}$ & 0 & 0 & 0 &$\mathbb{T}^2_{C}$\\
$e^c$ & $\mathbf{1}$ & $\mathbf{1}$ & $\mathbf{1}$ & 0 & 0 & \!$-6$\! &$\mathbb{T}^2_{C}$\\
$\mu^c$ & $\mathbf{1}$ & $\mathbf{1}$ & $\mathbf{1}$ & 0 & 0 & \!$-4$\! &$\mathbb{T}^2_{C}$\\
$\tau^c$ & $\mathbf{1}$ & $\mathbf{1}$ & $\mathbf{1}$ & 0 & 0 & \!$-2$\! &$\mathbb{T}^2_{C}$\\
$N_a^c$ & $\mathbf{1}$ & $\mathbf{1}$ & $\mathbf{1}$ & 0 & $-4$ & 0 &$\mathbb{T}^2_{B}$\\
$N_s^c$ & $\mathbf{1}$ & $\mathbf{1}$ & $\mathbf{1}$ & $-2$ & 0& 0 &$\mathbb{T}^2_{A}$\\
\hline 
$\Phi_{BC}$ & $\mathbf{1}$ & $\mathbf{3}$ & $\mathbf{3}$ & 0 & 0 & 0 & Bulk \\
$\Phi_{AC}$ & $\mathbf{3}$ & $\mathbf{1}$ & $\mathbf{3}$ & 0 & 0 & 0 & Bulk \\
\hline \hline
\end{tabular}
\begin{tabular}{| l | c c c c c c|}
\hline \hline
Yuk/Mass &$S_4^A$ & $S_4^B$ & $S_4^C$ & \!$2k_A$\! & \!$2k_B$\! & \!$2k_C$\!\\
\hline \hline
$Y_e(\tau_3)$ & $\mathbf{1}$ & $\mathbf{1}$ & $\mathbf{3}$ & 0 & 0 & $6$ \\
$Y_\mu(\tau_3)$ & $\mathbf{1}$ & $\mathbf{1}$ & $\mathbf{3}$ & 0 & 0 & $4$ \\
$Y_\tau(\tau_3)$ & $\mathbf{1}$ & $\mathbf{1}$ & $\mathbf{3}$ & 0 & 0 & $2$ \\
$Y_a(\tau_2)$ & $\mathbf{1}$ & $\mathbf{3}$ & $\mathbf{1}$ & 0 & $4$ & 0  \\
$Y_s(\tau_1)$ & $\mathbf{3}$ & $\mathbf{1}$ & $\mathbf{1}$ & $2$ & 0 & 0\\\hline
$M_a(\tau_2)$ & $\mathbf{1}$ & $\mathbf{1}$ & $\mathbf{1}$ & 0 & $8$ & 0  \\
$M_s(\tau_1)$ & $\mathbf{1}$ & $\mathbf{1}$ & $\mathbf{1}$ & $4$ & 0 & 0 
\\
\hline \hline
\end{tabular}
\caption{Transformation properties of fields and modular forms (Yuk/Mass) under the modular symmetries
$S^{A,B,C}_4$ with modular weights $k_{A,B,C}$. The Higgs fields $H_{u,d}$ (not displayed) transform trivially under all the modular $S_4$ symmetries. The leptons $L\sim (2,-1/2)$, and $e^c,\mu^c,\tau^c\sim (1,1)$ have the usual SM $SU(2)_L\times U(1)_Y$ quantum numbers
and the right-handed neutrinos $N^c_{a,s}$ are SM singlets.
The Higgs $\Phi$ which break the three modular symmetries to their diagonal subgroup, live in the 10d bulk,
while the leptons live in the 2d subspaces as shown.
}
\label{ta:model}
\end{footnotesize}
\end{table}

The fields Table \ref{ta:model} are interacting extra dimensional fields whose profiles are described in the Appendix \ref{app:eds}. The low energy phenomenology is studied after compactification.
The resulting 4d superpotential is~\cite{deMedeirosVarzielas:2022fbw}, ignoring the dimensionless coupling coefficients,
\begin{eqnarray}\label{eq:superpot}
w_\ell &=&
\frac{1}{\Lambda}\left[L \Phi_{BC} Y_a N_a^c + L \Phi_{AC} Y_s N_s^c \right] H_u \nonumber \\
&&+ \left[ L Y_e e^c + L Y_\mu \mu^c + L Y_\tau \tau^c \right] H_d \\
&&+ \frac{1}{2} M_a N_a^c N_a^c + \frac{1}{2} M_s N_s^c N_s^c   \nonumber \, .
\end{eqnarray}
and the modular Yukawa forms are fixed by the moduli $\tau_1=i, \tau_2=i+2, \tau_3=\omega$ resulting in the alignments, using Tables \ref{ta:modfor} and \ref{ta:model}, ignoring the overall constants,
\begin{equation}
\begin{split}
Y_a&=(0,1,-1)^T,\\
Y_s&=(1,1+\sqrt{6},1-\sqrt{6})^T,\\
Y_\tau&=(0,1,0)^T,\\
Y_\mu&=(0,0,1)^T,\\
Y_e&=(1,0,0)^T.
\end{split}
\label{eq:modforms}
\end{equation}
 
 The $\Phi$ fields are assumed to obtain a diagonal VEV that breaks two modular symmetries into the diagonal one \cite{deMedeirosVarzielas:2022fbw}.

Hence, the charged-lepton mass matrix is simply given by
\begin{equation}
M_l = v_d 
\begin{pmatrix}
 \left(Y_e\right)_{1} & \left(Y_\mu\right)_{1} & \left(Y_\tau\right)_{1} \\
 \left(Y_e\right)_{3} & \left(Y_\mu\right)_{3} & \left(Y_\tau\right)_{3} \\
 \left(Y_e\right)_{2} & \left(Y_\mu\right)_{2} & \left(Y_\tau\right)_{2} 
\end{pmatrix},
\end{equation}
where $v_d$ stands for $\left<H_d\right>$, and we ignore the dimensionless coupling coefficients.

Plugging in the specific shapes of the modular forms given in Eq. \ref{eq:modforms}
we arrive at a diagonal charged-lepton mass matrix for $\tau_C=\omega$, including the dimensionless coupling coefficients:
\begin{equation}
M_l = v_d 
\begin{pmatrix}
 y_e & 0 & 0 \\
 0 & y_\mu & 0 \\
 0 & 0 & y_\tau 
\end{pmatrix}.
\end{equation}

The Dirac neutrino mass matrix is then given by:
\begin{equation}
M_D = v_u 
\begin{pmatrix}
\left(Y_a\right)_1 & \left(Y_s\right)_1 \\
\left(Y_a\right)_3 & \left(Y_s\right)_3 \\
\left(Y_a\right)_2& \left(Y_s\right)_2  
\end{pmatrix},
\end{equation}
where, as usual, $v_u$ denotes the $H_u$ VEV, and the $2\times3$ structure comes from the CSD with just two RH neutrinos.
We have ignored the dimensionless coupling coefficients.
Choosing specific stabilisers for the two remaining moduli fields, we can achieve a CSD(3.45) structure with 
$n=1-\sqrt6$:
\begin{equation}
\label{eq:DiracMat}
M_D = v_u 
\begin{pmatrix}
0&   b\\
a&  b\left(1- \sqrt{6}\right) \\
-a & b\left(1+\sqrt{6}\right) 
\end{pmatrix}.
\end{equation}

The type-I seesaw mechanism will lead to an effective mass matrix for the light neutrinos:
\begin{equation}
m_\nu = M_D \cdot M_R^{-1} \cdot M_D^T = v_u^2 
\begingroup
\setlength\arraycolsep{15pt}
\begin{pmatrix}  
\dfrac{b^2}{M_s} & \dfrac{b^2 n}{M_s} &  \dfrac{b^2(2-n)}{M_s} \\[12pt]
. & \dfrac{a^2}{M_a} + \dfrac{b^2 n^2}{M_s} & -\dfrac{a^2}{M_a} + \dfrac{b^2n(2-n)}{M_s} \\[12pt] 
. & . & \dfrac{a^2}{M_a} + \dfrac{b^2(2-n)^2}{M_s}
\end{pmatrix},
\endgroup
\label{eq:mnu_mee}
\end{equation}
where $n=1-\sqrt{6} \approx -1.45$. This can be rewritten in terms of 3 independent physical parameters
\begin{equation}
m_\nu=m_a\left(\begin{matrix}
0 & 0& 0& \\
0 & 1 & -1\\
0 & -1 & 1
\end{matrix}\right)+m_be^{i\eta}\left(\begin{matrix}
1 & n& 2-n& \\
n & n^2 & n(2-n)\\
2-n & n(2-n) & (2-n)^2
\end{matrix}\right),
\end{equation}
where
\begin{equation}
m_a=\left|\frac{v_u^2 a^2}{M_a}\right|,\ \ \ m_b=\left|\frac{v_u^2 b^2}{M_s}\right|,\ \ \ \rho={\rm Arg}\left(\frac{a^2}{b^2}\right).
\end{equation}
Therefore the model has only these 3 parameters for the whole neutrino sector.

\subsection{Eclectic symmetry with the remnant $S_4$}

As discussed in Sec. \ref{sec:rems4}, the remnant symmetry $S_4$ acts on the branes and in the previous model, it has been identified with the modular symmetry $S_4^C$.
Therefore we have built a model whose flavour structure is completely defined by the modular symmetries $S_4^A\times S_4^B\times S_4^C$. However, it is known that having purely modular symmetries to define the flavour structure complicates the K\"ahler \cite{Chen:2019ewa}.
 
 In our setup, the only modular multiplet is the lepton doublet $L$. In general, the  minimal K\"ahler potential for a superfield $L$ would be a single term $\mathcal{K}= L\bar{L}$. However as $L$ is modular form, the K\"ahler potential is enhanced to include terms
 \begin{equation}
 \mathcal{K}=L\bar{L}+\sum_k a_k(Y_k \overline{Y_k})_\textbf{1} (L\overline{L})_\textbf{1}+\sum_k b_k(Y_k L)_\textbf{1} (\overline{Y_k L})_\textbf{1},
 \end{equation}
 where the sum is over all available modular forms and the fields inside a parenthesis $()_\textbf{1}$ are contracted into a modular symmetry singlet and the $a_k,b_k$ are arbitrary dimensionless constants. The $b_k$ terms appear only the case where $L$ is something larger than a singlet, like in our model, which is a triplet. In that case the sum is done over the three different modular forms from Table \ref{ta:modfor}, depending on the chosen $\tau$.
 
 The $a_k$ terms can be absorbed as an overall normalization of the field $L$, therefore they are not relevant. The $b_k$ terms are absorbed by the normalization of each component of the field $L$ and therefore introducing parameters that change the flavour structure given from the superpotential.
 
 In our model, the only nontrivial modular form is the lepton doublet $L$ which will have 3 $b_k$ K\"ahler terms. These parameters will affect the charged lepton mass matrix and can be reabsorbed in the definition of $y_{e,\mu,\tau}$, therefore preserving the same flavour structure there. However these parameters also affect the normalization of each left handed neutrino independently, introducing these 3 extra free parameters to the neutrino mass matrix and therefore reducing the predictiveness of the model. In general, for all modular symmetry models, it is assumed that these parameters are negligible, although they not necessarily need be.

One could avoid the presence of the unwanted terms by enhancing the modular symmetry by adding a standard flavour symmetry, such that the undesired K\"ahler terms are forbidden by the standard flavour symmetry \cite{Chen:2021prl}.
Relating flavor symmetries and modular symmetries are called eclectic symmetries \cite{Nilles:2020nnc,Nilles:2020kgo,Ding:2023ynd}.

As an alternative model to the one presented in the previous section, we could use the remnant symmetry as a standard flavour symmetry, as described in Sec \ref{sec:rems4}. This way the $S_4^C$ becomes standard flavour symmetry while $S_4^{A,B}$ remain as modular symmetries, thus having a trivial (where they all commute) eclectic symmetry.
With this assumption the lepton doublet is no longer a modular form and there are no $b_k$ K\"ahler terms. The model would require an extra $\mathbb{Z}_3$ shaping symmetry which would differentiate the three charged lepton singlets. Furthermore the modular forms $Y_{e,\mu,\tau}$ would not be available and they would have to be replaced by 3 flavon $S_4$ triplets $\phi_{e,\mu,\tau}$ whose VEV has the same desired alignments. This could be easily achieved through the orbifold boundary conditions and a very simple alignment superpotential \cite{deAnda:2018oik}.
With these changes, the flavour structure and all the phenomenological implications would be exactly the same as the model described in the previous subsection.
Thus the same flavour structure CSD($1-\sqrt{6}$) can be achieved easily through modular or eclectic $S_4^3$ symmetry.

\subsection{Numerical Fit}

\begin{table}[h]
\centering
  \begin{footnotesize}
    \begin{tabular}{c|l|cc|cc}
      \hline\hline
      &
      & \multicolumn{2}{c|}{without SK atmospheric data}
      & \multicolumn{2}{c}{with SK atmospheric data}
      \\
      \cline{3-6}
      && NuFit $\pm 1\sigma$ & Model
      & NuFit $\pm 1\sigma$ & Model
      \\
      \cline{2-6}
      \rule{0pt}{4mm}\ignorespaces
      & $\theta_{12}/^\circ$
      & $33.41_{-0.72}^{+0.75}$ & 34.34
      & $33.41_{-0.72}^{+0.75}$ & 34.30
      \\[3mm]
      & $\theta_{23}/^\circ$
      & $49.1_{-1.3}^{+1.0}$ & 48.31
      & $42.2_{-0.9}^{+1.1}$ & 46.98
      \\[3mm]
      & $\theta_{13}/^\circ$
      & $8.54_{-0.12}^{+0.11}$ & 8.54
      & $8.58_{-0.11}^{+0.11}$ & 8.75
      \\[3mm]
      & $\delta/^\circ$
      & $197_{-25}^{+42}$ & 284
      & $232_{-26}^{+36}$ & 278
      \\[3mm]
      & $\dfrac{\Dmq_{21}}{10^{-5}~\eVq}$
      & $7.41_{-0.20}^{+0.21}$ & 7.42
      & $7.41_{-0.20}^{+0.21}$ & 7.13
      \\[3mm]
      & $\dfrac{\Dmq_{3\ell}}{10^{-3}~\eVq}$
      & $+2.511_{-0.021}^{+0.028}$ & 2.510
      & $+2.507_{-0.027}^{+0.026}$ & 2.520\\[3mm]
      \hline \hline &&&& \\[-3mm]
      & $\dfrac{m_a}{10^{-3}~{\rm eV}}$
      &  & $31.47$
      &  & 30.50\\[3mm]
      & $\dfrac{m_b}{10^{-3}~{\rm eV}}$
      &  & $2.28$
      &  & 2.32\\[3mm]
      & $\eta/\pi$
      & &  $1.24$
      &  & $1.26$\\[3mm]
      \hline
        & $\chi^2$
      & & $6.3$
      &  & 26.61\\
      \hline\hline
    \end{tabular}
  \end{footnotesize}
  \caption{Normal Ordering NuFit~5.2 values~\cite{Esteban:2020cvm, nufit} for the neutrino observables, and the best fit point from the model. The best fit is for NuFit data without SK atmospheric data where the atmospheric angle $\theta_{23}$ is in the second octant,
  as preferred by the model. }
  \label{ta:NuFit52}
\end{table}

With the CSD($1-\sqrt{6}$) structure, we can achieve the fits shown in Table \ref{ta:NuFit52}.
Note that in both best fits, there is a unique physical phase $\eta\approx 5\pi/4$ which could point to a geometrical origin.

To quantify how good the fit is we use 
\begin{equation}
\chi^2=\sum_i\left(\frac{x_i^{\rm exp}-x_i^{\rm model}}{\sigma_i^{\rm exp}}\right)^2,
\end{equation}
where it is summed over all 6 experimental neutrino values $(\theta_{12},\theta_{13},\theta_{23},\delta,\Dmq_{21},\Dmq_{3\ell})$. The model fits these 6 observables plus the lightest neutrino mass (which is zero), the Majorana phases (where one is unphysical) which determine neutrinoless beta decay parameter which is just equal to the $(1,1)$ element of the neutrino mass matrix,
$m_{ee}=m_b$ \cite{deMedeirosVarzielas:2022fbw}. Note that the 6 experimentally constrained observables are being fit with only 3 real parameters $(m_a,m_b,\eta)$, which is a non-trivial achievement. Overall these 3 parameters are predicting 9 neutrino observables, which shows that the model is highly predictive.
In particular the model requires a normal neutrino mass squared ordering with the lightest neutrino being massless,
and predicts the atmospheric angle to be in the second octant, 
$\theta_{23}\approx 48^\circ$, with close to maximal leptonic CP violation, $\delta\approx 280^\circ$.

\section{Conclusions}
\label{sec:conclusions}

In recent years modular symmetries have been applied to flavour models in bottom-up approaches, where finite modular groups 
$\Gamma_N$ may result from the quotient group of the modular symmetry by its principal congruence subgroup of 
level $N$, where for example $N=4$ corresponds to $S_4$. In such approaches the role of the flavon field is played by a complex modulus field $\tau$ in orbifold models with two extra dimensions.

In this paper we have discussed modular symmetry models arising from bottom-up orbifold constructions.
The simplest example in 6d involves the orbifold $\mathbb{T}^2/\mathbb{Z}_N$ with a single torus defined by one complex coordinate 
$z$ and a single modulus field $\tau$, playing the role of a flavon transforming under a finite modular symmetry.
More generally we have considered bottom-up orbifolds in 10d, where the 6 extra dimensions are 
factorisable into 3 tori, each defined by one complex coordinate $z_i$ and involving the three moduli fields $\tau_1, \tau_2, \tau_3$ transforming under three independent finite modular groups.
Assuming supersymmetry, consistent with the holomorphicity requirement, 
we consider all the orbifolds of the form $(\mathbb{T}^2)^3/(\mathbb{Z}_N\times\mathbb{Z}_M)$,
and list all the available orbifolds, which have fixed values of the moduli fields (up to an integer).
The key advantage of such 10d orbifold models over 4d models is that the values of the moduli are not completely free but are 
constrained by geometry and symmetry.

To illustrate the approach we have shown how a recently proposed littlest modular seesaw model with $S_4^3$ modular symmetry could result from such an orbifold construction. We have shown how this model may arise from 
an $(\mathbb{T}^2)^3/(\mathbb{Z}_4\times\mathbb{Z}_2)$ orbifold 
with $\tau_1=i,\ \tau_2=i+2$ being fixed by the geometry of the $\mathbb{Z}_4$ orbifold,
while $\tau_3=\omega$ is determined by imposing a further remnant $S_4$ flavour symmetry,
commuting with the three $S_4$ modular symmetries. 
The $\tau_1=i$ leads to an alignment $(1,1+\sqrt{6},1-\sqrt{6})$ and
CSD$(n)$ with $n=1-\sqrt{6}$, where the atmospheric angle $\theta_{23}$ is restricted to lie in the second octant. The $\chi^2$ fit shows that this is a highly predictive and successful description of all neutrino phenomenology,
with two real parameters describing all neutrino mixing and mass ratios.

An alternative case $n=1+\sqrt{6}$, which prefers the atmospheric angle $\theta_{23}$ to be in the first octant, 
which was possible in the 4d model \cite{deMedeirosVarzielas:2022fbw}, is not allowed here since it corresponds to an alignment 
$(1,1-\sqrt{6},1+\sqrt{6})$ and a stabilizer $\tau_1=(-8+i)/13$ which is not achievable in the 10d orbifold model considered here. Whereas in the 4d model \cite{deMedeirosVarzielas:2022fbw}, the fixed points were selected in an {\it ad hoc} way, in the 10d orbifold model considered here the values of $\tau_{1,2}$ are fixed by the geometry to be equal to $i$ (up to an integer).

In the modular symmetry model considered here, 
the remnant $S_4$ plays no role apart from fixing $\tau_3=\omega$. However in an alternative orbifold model,
the combination of flavour symmetry and modular symmetry could be used to control the corrections to the 
K\"ahler potential, as in top-down eclectic flavour symmetry. This would reintroduce flavons, and lead to a more complicated
model which is beyond the scope of the main discussion here, although we have briefly sketched the consequences.

Finally we note that the bottom-up approach to modular symmetry from orbifolds followed here 
can readily be extended to GUTs, with up to three moduli groups and moduli fields.
One could similarly include a remnant flavour symmetry, leading to a bottom-up version of the ecletic flavour symmetry in orbifold GUTs.

\section*{Acknowledgements}
SFK acknowledges the STFC Consolidated Grant ST/L000296/1 and the European Union's Horizon 2020 Research and Innovation programme under Marie Sklodowska-Curie grant agreement HIDDeN European ITN project (H2020-MSCA-ITN-2019//860881-HIDDeN).
SFK would also like to thank CERN for its hospitality.

\appendix

\section{Extra dimensional fields}
\label{app:eds}

We assume $\mathcal{N}=1$ Super Yang-Mills in 10 dimensions where the gauge fields propagate in the bulk as well as the  superfields $\Phi, H_{u,d}$. The 6 extra dimensions are compactified into three factorisable flat tori.

Let us work out the profiles for the first torus which must fulfill
\begin{equation}
    z_1\sim z_1+2\pi R_1,\ \ \ z_1\sim z_1+ \tau_1 2\pi R_1,
\end{equation}
where the extra dimensional coordinate
\begin{equation}
    z_1=x_5+ix_6=y_5+\tau_1 y_6,
\end{equation}
where $y_{5,6}\in [0,2\pi R_1]$ and
\begin{equation}
    \begin{array}{ll}
    x_5=\Re(z_1), & x_6=\Im(z_1),\\
    y_5=\Re(z_1)-\frac{\Re(\tau_1)}{\Im(\tau_1)}\Im(z_1),& y_6=\frac{\Im(z_1)}{\Im(\tau_1)},
    \end{array}
\end{equation}
from which the ED profile is determined to be
\begin{equation}
    \xi_{n_1 m_1}(z_1)\sim e^{i(n_1y_5+m_1y_6)/R_1}=e^{i2\Re[M_{m_1 n_1} z_1]/R_1},
\end{equation}
where $n_1,m_2$ are integers. These are orthogonal functions
\cite{Nilse:2007zz}
\begin{equation}
    \int dz dz^* \xi_{n_1m_1}(z_1)\xi_{n'_1 m'_1}(z_1)\sim \delta^{n_1}_{n'_1}\delta^{m_1}_{m'_1},
\end{equation}
and they are translation eigenfunctions 
\begin{equation}
\begin{split}
    \partial_{z_1} \xi_{n_1 m_1}(z_1)=&M_{m_1 n_1}\xi_{n_1 m_1}(z_1).
    \end{split}
\end{equation}
This can be repeated for each of the 3 tori.

The 10 dimensional fields can be decomposed into a Fourier series
\begin{equation}
\Phi(x,z_1,z_2,z_3)=\sum_{n_1,n_2,n_3,m_1,m_2,m_3}\Phi_{n_1 n_2 n_3 m_1 m_2 m_3}(x) \xi_{n_1 m_1}(z_1) \xi_{n_2 m_2}(z_2)  \xi_{n_3 m_3}(z_3),
\label{10d}
\end{equation}
where $n_i,m_i$ are integers and only the massless zero mode $\Phi_{000000}(x)$ is relevant at low energies.
The massless modes of 10d bulk fields, which in the present model include the Higgs fields, will appear in  the 
effective superpotential in Eq. \ref{eq:superpot}.

All of the SM fermions belong to 6d superfields located in different tori $\mathbb{T}^2_{A,B,C}$ which are 6 dimensional subspaces where $\mathcal{N}=1$ 6d SUSY is preserved. These superfields containing the SM fermions, for example the $L$, can be written as 10d superfields and decomposed as
\begin{equation}
L(x,z_1,z_2,z_3)=\sum_{n_3,m_3}L_{n_3 m_3}(x)\xi_{n_3 m_3}(z_3)\delta(z_1)\delta(z_1^*)\delta(z_2)\delta(z_2^*).
\label{6d}
\end{equation}
Due to the Dirac delta functions, this field is effectively a 6d field and transforms under the modular symmetry appropriate to that 6d subspace~\cite{Ferrara:1989bc,Ferrara:1989qb}.
As in Eq.~\ref{10d}, it may be expanded in a Fourier series, with the zero modes appearing in  the 
effective superpotential in Eq. \ref{eq:superpot}.

Note that the boundary conditions are chosen such that all the SM superfields have a single zero mode and 
$\mathcal{N}=1$ 4d SUSY is preserved after compactification.
Writing all the 6d fermion fields in Table \ref{ta:model} as in Eq.~\ref{6d}, then expanding these fields,
together with the 10d Higgs fields as a Forurier series in Eq.~\ref{10d}, one can identify the most relevant 10d interaction terms which, after compactification, involves the zero modes which appear in the resulting
effective 4d superpotential in Eq. \ref{eq:superpot}.

\end{document}